\documentclass[11pt,twoside]{article}
\usepackage{./asp2014}

\aspSuppressVolSlug
\resetcounters

\begin{document}

\title{Metallicity Distribution of Galactic Halo Field RR~Lyr\ae{}, and the Effect of Metallicity on their Pulsation Properties}

% Authors list
\author{Marengo, M.,$^1$ Mullen, J.,$^1$ Neeley, J. R.,$^2$ Fabrizio, M.,$^{3,4}$ Marrese, P. M.,$^{3,4}$ Bono, G.,$^{3,5}$ Braga, V. F.,$^{3,4}$ Magurno, D.,$^5$ Crestani, J.,$^5$ Fiorentino, G.,$^3$ Monelli, M.,$^6$ Chaboyer, B.,$^7$ Gilligan, C. K.,$^7$ Dall'Ora, M.,$^8$ Mart\'inez-V\'azquez, C. E.,$^9$ Th\'evenin, F.$^{10}$ and Matsunaga, N.$^{11}$
\affil{$^1$Iowa State University, USA; \email{mmarengo@iastate.edu}}
\affil{$^2$Florida Atlantic University, USA}
\affil{$^3$INAF - Osservatorio Astronomico di Roma, Italy}
\affil{$^4$ASI - Space Science Data Center, Italy}
\affil{$^5$Universit\'a di Roma Tor Vergata, Italy}
\affil{$^6$Instituto de Astrof\'isica de Canarias, Spain}
\affil{$^7$Dartmouth College, USA}
\affil{$^8$INAF - Osservatorio Astronomico di Capodimonte, Italy}
\affil{$^9$Cerro Tololo Inter-American Observatory, Chile}
\affil{$^{10}$ Observatoire de la C\^ote d'Azur, France}
\affil{$^{11}$ University of Tokyo, Japan}
}

% This section is for ADS Processing.  There must be one line per author.
\paperauthor{Marengo, M.}{mmarengo@iastate.edu}{0000-0001-9910-9230}{Iowa State University}{Physics and Astronomy}{Ames}{IA}{50010}{USA}
\paperauthor{Mullen, J.}{jpmullen@iastate.edu}{0000-0002-5138-3246}{Iowa State University}{Physics and Astronomy}{Ames}{IA}{50010}{USA}
\paperauthor{Neeley, J. R.}{neeleyj@fau.edu}{0000-0002-8894-836X}{Florida Atlantic University}{}{Boca Raton}{FL}{}{USA}
\paperauthor{Fabrizio, M.}{michele.fabrizio@ssdc.asi.it}{0000-0001-5829-111X}{INAF - Osservatorio Astronomico di Roma}{}{Monte Porzio Catone}{}{}{Italy}
\paperauthor{Marrese, P. M.}{paola.marrese@ssdc.asi.it}{0000-0002-8162-3810}{INAF - Osservatorio Astronomico di Roma}{}{Monte Porzio Catone}{}{}{Italy}
\paperauthor{Bono, G.}{bono@roma2.infn.it}{0000-0002-4896-8841}{Universita degli Studi di Roma Tor Vergata}{}{Roma}{}{}{Italy}
\paperauthor{Braga, V.}{vittorio.braga@roma2.infn.it}{0000-0001-7511-2830}{Universita degli Studi di Roma Tor Vergata}{}{Roma}{}{}{Italy}
\paperauthor{Magurno, D.}{davide.magurno@roma2.infn.it}{0000-0001-5479-5062}{Universita degli Studi di Roma Tor Vergata}{}{Roma}{}{}{Italy}
\paperauthor{Crestani, J.}{crestani.juli@gmail.com}{0000-0002-7717-9227}{Universita degli Studi di Roma Tor Vergata}{}{Roma}{}{}{Italy}
\paperauthor{Fiorentino, G.}{giuliana.fiorentino@inaf.it}{0000-0003-0376-6928}{INAF - Osservatorio Astronomico di Roma}{}{Monte Porzio Catone}{}{}{Italy}
\paperauthor{Monelli, M.}{monelli@iac.es}{0000-0001-5292-6380}{Instituto de Astrofisica de Canarias}{}{La Laguna}{Tenerife}{}{Spain}
\paperauthor{Chaboyer, B.}{brian.chaboyer@dartmouth.edu}{0000-0003-3096-4161}{Dartmouth College}{}{Hanover}{NH}{}{USA}
\paperauthor{Gilligan, C. K.}{christina.k.gilligan.gr@dartmouth.edu}{0000-0003-4510-0964}{Dartmouth College}{}{Hanover}{NH}{}{USA}
\paperauthor{Dall'Ora, M.}{allora@na.astro.it}{0000-0001-8209-0449}{INAF - Osservatorio Astronomico di Capodimonte}{}{Napoli}{}{}{Italy}
\paperauthor{Martinez-Vazquez, C. E.}{cmartinez@ctio.noao.edu}{0000-0002-9144-7726}{Cerro Tololo Inter-American Observatory}{}{La Serena}{}{}{Chile}
\paperauthor{Thevenin, F.}{frederic.thevenin@oca.eu}{}{Observatoire de la Cote d'Azur}{}{Nice}{}{}{France}
\paperauthor{Matsunaga, N.}{matsunaga@astron.s.u-tokyo.ac.jp}{}{The University of Tokyo}{Department of Astronomy}{Tokyo}{}{113-0033}{Japan}

\begin{abstract}
We present our analysis of a large sample (over 150k) of candidate Galactic RR Lyrae (RRL) stars for which we derived high quality photometry at UV, optical and infrared wavelengths, using data from publicly available surveys. For a sub-sample of these stars ($\sim$  2,400 fundamental mode field RRLs) we have measured their individual metallicity using the $\Delta$S method, resulting in the largest and most homogeneous spectroscopic data set collected for RRLs. We use this sample to study the metallicity distribution in the Galactic Halo, including the long-standing problem of the Oosterhoff dichotomy among Galactic globular clusters. We also analyze the dependence of their pulsation properties, and in particular the shape of their infrared light curves, from their [Fe/H] abundance.
\end{abstract}

\section{Introduction}\label{sec-1}

For the past 70 years RR Lyr\ae{} variables (RRLs) have been widely used as tracers of old stellar populations in the Milky Way and Local Group galaxies, thanks to their luminosity-metallicity relation (see e.g.  \citealt{2015pust.book.....C} for a review). However, a large ($> 5\%)$ intrinsic scatter in this relation at visible wavelengths, mainly due to evolutionary effects and temperature dependence, has limited the precision of these stars as distance indicators. This has changed with the recent calibration of theoretical (e.g. \citealt{2015ApJ...808...50M, 2017ApJ...841...84N}) and observational (see e.g. \citealt{2013MNRAS.435.3206D, 2018MNRAS.481.1195M, 2019MNRAS.490.4254N}) period-Wesenheit and period-luminosity relations (in the optical and infrared, respectively), revealing the potential of these stars as high precision standard candles. The same studies have however also highlighted how metallicity plays an important role in the absolute brightness of these stars, and should be taken into account when calibrating their distances, effectively requiring period-luminosity-metallicity (PLZ) and period-Wesenheit-metallicity (PWZ) relations.

Currently, the largest source of uncertainty in the calibration of RRLs PLZ and PWZ relations is related to systematics in the Gaia parallaxes of RRLs calibrators (see e.g. \citealt{2019MNRAS.490.4254N}). These issues are expected to be mitigated in forthcoming Gaia releases, at which point the main obstacle to obtain precise RRLs distances will remain the difficulty of obtaining reliable measurements of RRLs metallicity. Currently available catalogs (see e.g. \citealt{2013MNRAS.435.3206D} for one of the largest, recent compilations) tend to list metallicities measured with heterogeneous methods calibrated on the basis of different metallicity scales. While spectroscopic determinations using high resolution spectra and a multitude of lines tend to be more accurate, they are also more expensive to acquire, and as such limited to small samples of stars. Photometric methods, and methods relying on low resolution spectra, are more easily applied to large samples of RRLs, but often lack enough precision for an accurate calibration of PLZ and PWZ relations, and hinder studies using RRLs are probes of the metallicity distribution in galaxies and globular clusters.

To address this issue, we have compiled a large catalog of field RRLs, with multi-wavelength photometry already available in the literature, matched with parallaxes from Gaia DR2 \citep{2019ApJ...882..169F}. For a subsample of these stars we have derived their [Fe/H] abundance on a homogeneous scale using publicly available low resolution spectra. The process in which this catalog was created is summarised in section~\ref{sec-2}. In section~\ref{sec-3} we use our metallicity sample to study the [Fe/H] abundance distribution in the Galactic Halo, with particular emphasis to the  Oosterhoff dichotomy in Globular Clusters. Finally, in section~\ref{sec-4} we study the effect of metallicity on the pulsation properties of the stars, in particular the Fourier parameters of their infrared light curves.

\section{A large catalog of homogeneous RRLs metallicities}\label{sec-2}

The procedure followed to create our RRL catalog  is summarised in \citet{2019ApJ...882..169F}, and will be described in detail in two forthcoming papers (Bono et al., Marinoni et al. in preparation). The catalog, which contains photometry in the UV, visible and infrared wavelengths for more than 150,000 RRab (fundamental) and RRc (first overtone) unique variables, has been compiled by combining data from a variety of publicly available surveys. These data sets have been cross-matched with the Gaia DR2 survey using an algorithm specifically developed for sparse catalogs \citep{2019A&A...621A.144M}. 

For our metallicity study we derived an homogeneous set of [Fe/H] abundances for a subsample of these stars ($\sim $2,400 RRab variables in the field) using low resolution ($R \sim$ 2,000) spectra mainly collected as part of the SDSS DR12 SEGUE survey \citep{2015ApJS..219...12A}. Our metallicities were measured with the $\Delta S$ method \citep{1959ApJ...130..507P, 1994AJ....108.1016L}, based on the ratio between the equivalent widths of Ca \textsc{ii} K, H$\beta$, H$\gamma$ and H$\delta$ lines. The method was re-calibrated and validated using a subsample of these stars for which we obtained high precision abundances using high resolution ($R >$ 20,000) spectra (mainly from \citealt{2018ApJ...864...57M}), complemented with metallicities derived by \citet{2013MNRAS.435.3206D}. Our [Fe/H] abundances have been carefully matched to the \citet{1994AJ....108.1016L} metallicity scale and have an estimated accuracy of $\sigma_{[Fe/H]} \simeq 0.29$~dex. We believe this is the largest available sample of homogeneous metallicities for field RRab variables in the Galactic Halo; a similar sample focused on RRc pulsators is currently in preparation.

\articlefigure[width=1.0\textwidth]{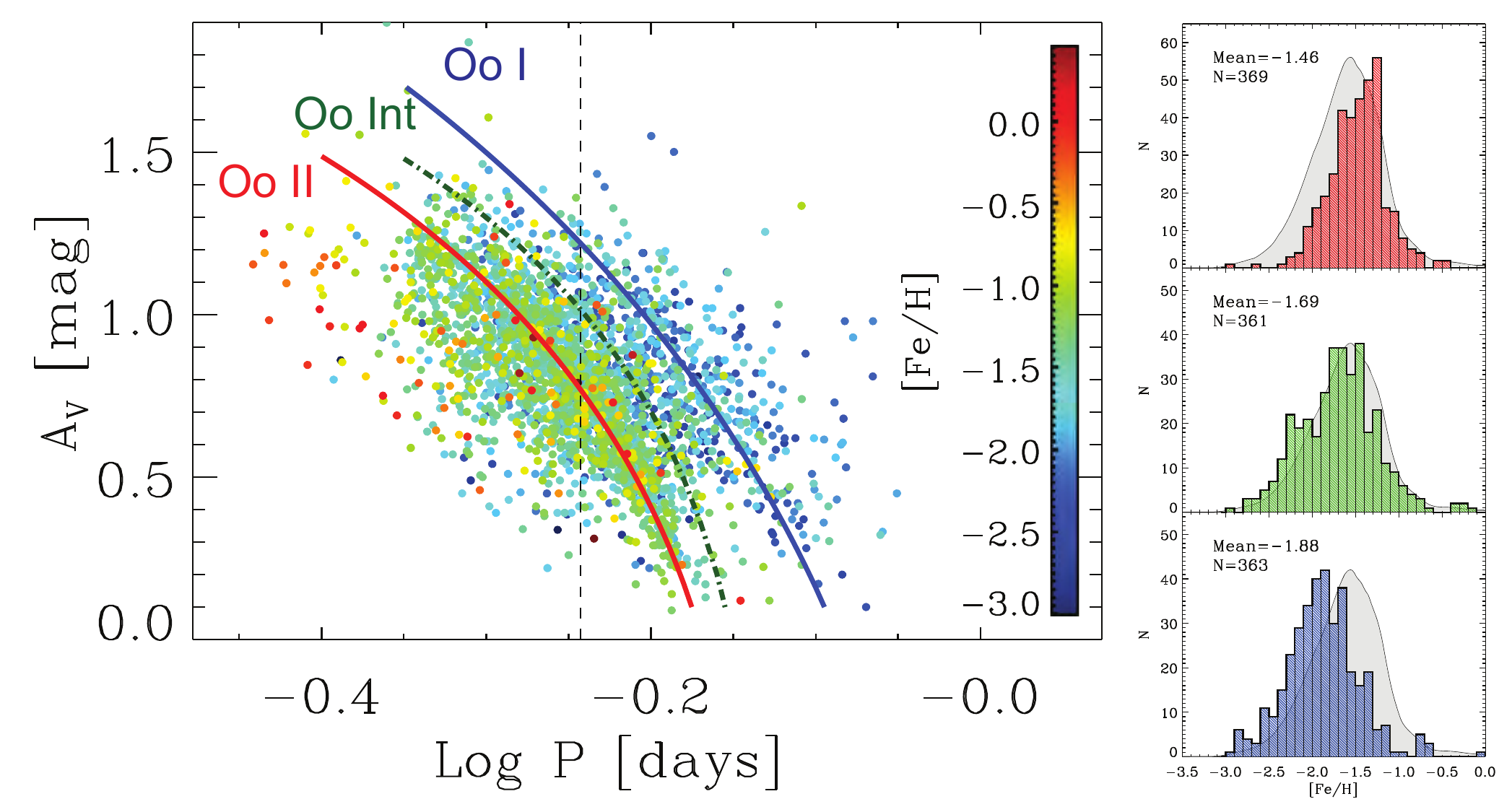}{fig1}{\emph{Left:} Bailey diagram of our metallicity sample, color-coded on the basis of their [Fe/H] abundance. The blue, red and green lines mark the loci we have determined for the OoI, OoII and OoInt sequences. \emph{Right:} metallicity distribution of the variables along the three Oosterhoff sequences, compared to the overall metallicity distribution of the sample (in grey). Adapted from \citet{2019ApJ...882..169F}.}

\section{[Fe/H] abundance distribution in Galactic Halo RRLs}\label{sec-3}

Galactic globular clusters can be split according to the so-called Oosterhoff dichotomy \citep{1939Obs....62..104O}, on the basis of the average period of their RRLs: Oosterhoff type I clusters (OoI, $\langle P \rangle \sim 0.56$~days), and Oosterhoff type II clusters (OoII, $\langle P \rangle \ga 0.66$~days). The same is not true for some Local Group galaxies and their globular clusters, whose RRLs fill the period gap, revealing a population of Oosterhoff intermediate (OoInt) systems. The lack of OoInt clusters in the Galaxy suggests that the Oosterhoff dichotomy is related to the environment (see e.g. \citealt{2015ApJ...798L..12F}). Metallicity certainly plays a role, since it has long been known that OoI clusters tend to be more metal rich than their OoII counterparts \citep{1955AJ.....60..317A, 1959MNRAS.119..134K}.

While the Oosterhoff dichotomy in the Galaxy has been well studied for clusters, population studies of field RRLs in the Halo have been lacking, mainly due to the difficulty of compiling large catalogs of field RRLs with homogeneous [Fe/H] abundances. The catalog we presented in \citet{2019ApJ...882..169F} is ideally suited for this work, by allowing a quantitative analysis of the fine structures in the period-amplitude diagram (Bailey diagram), and their relation with the metallicity of individual stars.

Figure~\ref{fig1} shows the Bailey diagram of our spectroscopic sample, color-coded on the basis of their $\Delta S$ metallicity. We identify two ridges in the period-amplitude distribution, corresponding to the loci expected for OoI and OoII populations. These are marked with the blue and red lines, respectively. We also find a valley between these two overdensities, associated to the OoInt population (in green). Analytical formulas for these three loci are provided in \citet{2019ApJ...882..169F}. The right panels in figure~\ref{fig1} instead show the [Fe/H] distribution of RRLs on narrow strips along the OoI, OoII and OoInt sequences, compared to the overall metallicity distribution of the sample (in grey). It is clear that \emph{even in the field}, OoI stars are more metal poor than OoII RRLs, with OoInt variables in-between. This trend is corroborated by the presence of a continuous and linear relation between [Fe/H] abundance and period, which we have calibrated with our data and is also presented in \citet{2019ApJ...882..169F}.

This result supports the hypothesis that the Oosterhoff dichotomy in globular clusters is mainly the consequence of the lack of Galactic metal-intermediate clusters hosting RRLs, confirming theoretical and empirical evidences brought forward in the literature \citep{1983MmSAI..54..141C, 1983MmSAI..54..335R}.

\articlefigure[width=1.0\textwidth]{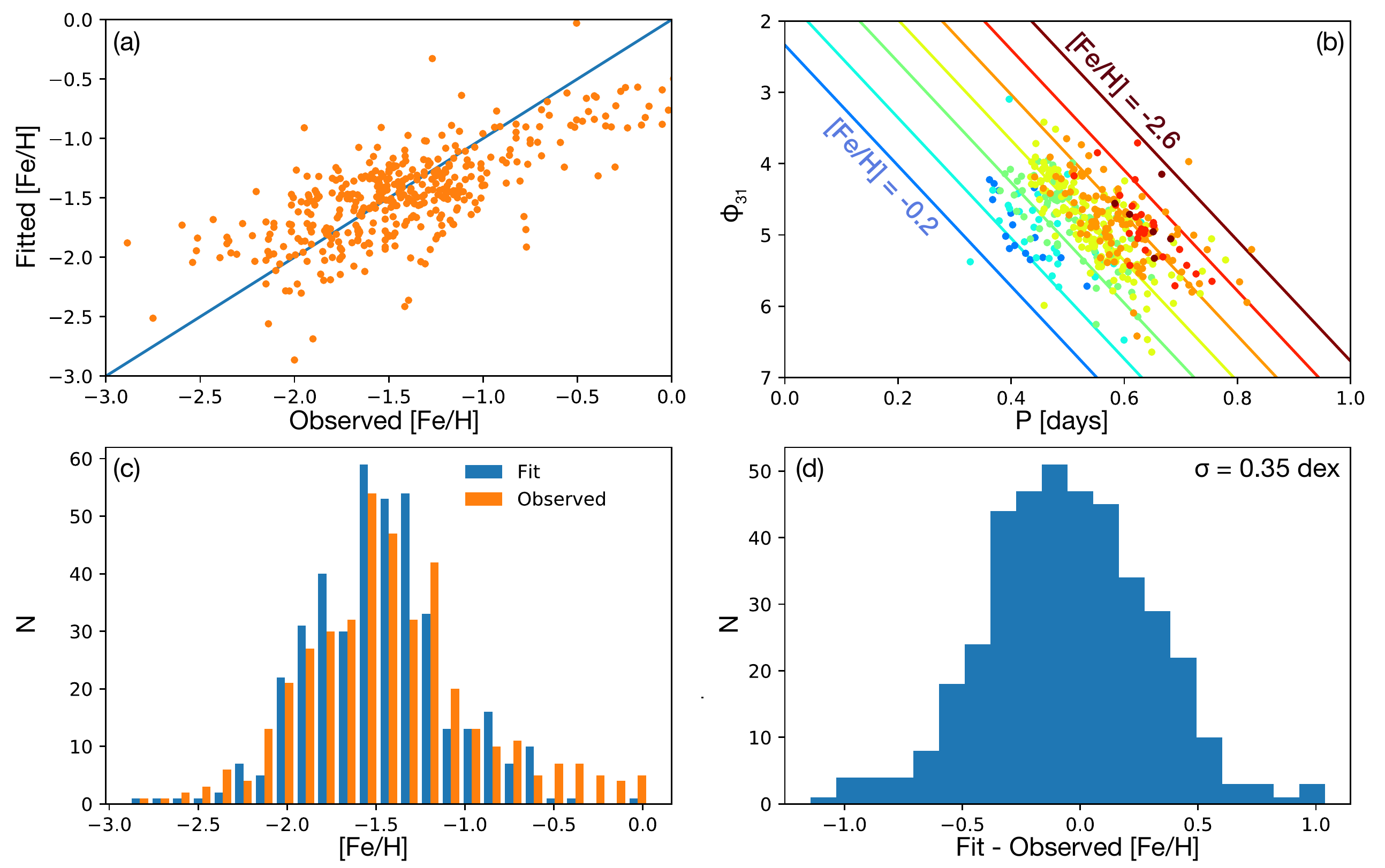}{fig2}{Metallicity-Period-$\phi_{31}$ relation for our sample stars: \emph{(a)} fitted vs. measured [Fe/H] abundance; \emph{(b)} distribution of our stars in the $\phi_{31}$ vs. $P$ plane, color coded by metallicity and compared with best fit lines calculated for different metallicities; \emph{(c)} comparison between best fit and measured [Fe/H] distribution; \emph{(d)} distribution of residuals between best fit and measured [Fe/H] abundances.}

\section{Effect of metallicity on the pulsation properties}\label{sec-4}

The analysis presented in section~\ref{sec-3} provides a clear and quantitative confirmation that the pulsation properties (e.g. period and amplitude) of RRab stars are affected by their [Fe/H] abundance. More in general, it is well known that the shape itself of the RRLs visible light curve is affected by metallicity \citep{1995A&A...293L..57K}, and that the [Fe/H] abundance can be estimated from a linear relation between period and Fourier parameters \citep{1996A&A...312..111J}, especially the phase $\phi_{31} = \phi_3 - i \phi_1$ parameter.

To improve on this analysis we have collected well-sampled light curves for the stars in our metallicity sample. Our current efforts cover the entire visible and infrared spectrum, taking advantage of publicly available surveys, as well as our own observations. As a test case we have extracted high quality NeoWISE light curves \citep{2014ApJ...792...30M} with 120 individual epochs for over 600 stars in the $W1$ (3.4~$\mu$m) and $W2$ (4.6~$\mu$m) bands, for over 600 stars in our catalog. The use of infrared bands is of particular interest for this analysis, since it allows us to test if the relation between metallicity, period and Fourier parameters still hold in a regime where the shape of the light curve is dominated by the changes in stellar radius, as opposed to changes in temperature, as in visible bands.

Figure~\ref{fig2} shows that the metallicity of our sample of RRab stars can be recovered using a bi-linear function of the period $P$ and $\phi_{31}$, with an accuracy $\sigma \simeq 0.35$~dex (compared to the 0.29~dex estimated accuracy for the $\Delta S$ method abundances). The fit works especially well for the intermediate range of metallicities (where we have the largest number of calibrators), but tends to overpredict the abundance of RRab variables for [Fe/H] $\ga -0.7$, where however the metallicity calibration of our sample is poorly constrained. Further work, and a better constrained calibration of the $\Delta S$ method in the high metallicity regime, is required to ascertain if this deviation is real. The result of this analysis, compared with similar relations derived for light curves collected in the visible, will be presented in a forthcoming publication.

\acknowledgements This work was partially supported by the National Science Foundation under grant No. AST1714534.

\end{document}